\documentstyle[aps,preprint]{revtex}
\begin{document}
\title{Anomalous flux quantization in a Hubbard ring \\
with correlated hopping.}
\author{Liliana Arrachea\thanks{%
Consejo Nacional de Investigaciones Cient\'{\i}ficas y T\'ecnicas} $^{\dag}$%
, A. A. Aligia and E. Gagliano $^{*}$.}
\address{Centro At\'{o}mico Bariloche and Instituto Balseiro \\
Comisi\'on Nacional de Energ\'{\i}a At\'{o}mica \\
8400 Bariloche, Argentina \\
$^{\dag}$ Permanent address: Departamento de F\'{\i}sica, Universidad\\
Nacional de La Plata, \\
1900 La Plata, Argentina.}
\maketitle

\begin{abstract}
We solve exactly a generalized Hubbard ring with twisted boundary
conditions. The magnitude of the nearest-neighbor hopping depends on the
occupations of the sites involved and the term which modifies the number of
doubly occupied sites $t_{AB}=0$. Although $\eta$-pairing states with
off-diagonal long-range order are part of the degenerate ground state, the
behavior of the energy as a function of the twist rules out
superconductivity in this limit. A small $t_{AB}$ breaks the degeneracy and
for moderate repulsive $U$ introduce superconducting correlations which lead
to ``anomalous'' flux quantization. 
\end{abstract}

\newpage
One of the most interesting problems of the physics of highly correlated
electronic systems is the characterization of the metallic, insulating and
superconducting phases, as well as the transitions among them. Kohn has
shown that the Drude weight $D_c$ is the adequate quantity to identify the
metal-insulator transition (MIT) \cite{kohn}, while Yang introduced the
concept of off-diagonal long-range order (ODLRO) to characterize the
superconducting nature of a metallic phase \cite{yang}. ODLRO in {\it all}
relevant low-energy eigenstates implies a periodicity of $h/2e$ in the free
energy as a function of a magnetic flux threading a system with annular
topology, which is referred to as ``anomalous'' flux quantization (AFQ) \cite
{bye,kohn2}. In other words, AFQ in the ground-state (GS) means $%
E(\Phi+\pi)=E(\Phi)$, where $E$ is the GS energy and $\Phi$ is the twist
angle. It is a necessary but  not sufficient condition  for
superconductivity \cite{note}. Since $D_c \sim \partial^2 E/\partial \Phi^2$%
, the function $E(\Phi)$ gives crucial information about the metallic and
superconducting character of the system \cite{kohn,kohn2,scala}.

Exactly solvable highly correlated models displaying a MIT or ODLRO are good
laboratories to investigate the nature of the MIT and electronic mechanisms
of superconductivity. The Bethe ansatz solution with twisted boundary
conditions (i.e. arbitrary flux) of the one dimensional (1D) Hubbard model
\cite{sha}, allowed to apply Kohn's ideas to the MIT in this model \cite
{sha,kaw}. 
Very few exact results exist, for electronic models exhibiting
superconductivity (or dominant superconducting correlations at long
distances in 1D). Several of them are related with the so called $\eta$%
-pairing mechanism, which allows to construct eigenstates with ODLRO \cite
{yan2,li1,boe,sch}. In particular, the widely studied \cite{note0} Hubbard
model  with correlated hopping,
\begin{eqnarray}
H= H_U\;+\;H_t &=& U\sum_i n_{i\uparrow} n_{i\downarrow}\; +\;
\sum_{<ij>\sigma}(c^{\dagger}_{i \bar \sigma} c_{j \bar \sigma} +h.c) \{
t_{AA}\;(1-n_{i\sigma}) (1-n_{j\sigma})+  \nonumber \\
& & t_{BB}\;n_{i\sigma} n_{j\sigma} + t_{AB}\;[n_{i\sigma}(1-n_{j\sigma}) +
n_{j\sigma}(1-n_{i\sigma})] \},  \label{1}
\end{eqnarray}
has been exactly solved recently in 1D in the limit $t_{AB}=0$ for open \cite
{li1} and periodic \cite{boe,sch} boundary conditions. It has been shown
that, due to an SU(2) pseudospin symmetry, $\eta$-pairing  states with ODLRO
are part of the {\em degenerate} GS for moderate on-site repulsion $U$ and
arbitrary band filling. Unfortunately, the function $E(\Phi)$ has not been
obtained and then, the AFQ and $D_c$ were not studied. Our main interest in
this study is motivated by the following two facts: first, the
superconducting character of the degenerate GS is not obvious, even when
states with ODLRO are part of the GS manifold.
Second, the GS was found to be a Mott insulator for $%
U>U_{MI}=2D(|t_{AA}|+|t_{BB}|)$ at half-filling (density of particles $n=1$%
), in a simple cubic lattice in D dimensions, with a MIT for D$>1$ \cite
{ali,str}. Strictly in D$=1$, however, we find that for $n=1$, $D_c=0 \;
\forall U$, {\em in spite of a vanishing charge gap} for $U<U_{MI}$. The
possibility of an insulating phase with this feature was first remarked by
Kohn and we think that this is, to our knowledge, the first non trivial
realization of that kind of insulators.

In this Letter, we solve exactly the model (\ref{1}) with $t_{AB}=0$, for
twisted boundary conditions $\Phi_{\uparrow}\; (\Phi_{\downarrow})$ for spin
up (down) fermions. This allows us to calculate the Drude weight $D_c$ and
the spin stiffness $D_s$ and to discuss the nature of the MIT as $%
n\rightarrow 1$. The behavior of $E(\Phi_{\uparrow}, \Phi_{\downarrow})$,
rules out superconductivity in the model for $t_{AB}=0$, at least in the
one-dimensional case we study here. In addition, we go one step further and
show how the GS degeneracy is broken in favor of a state with dominant
superconducting correlations  when a finite $t_{AB}$ is allowed, for
moderate repulsive $U$.

We first consider the Hamiltonian (\ref{1}) with $-t_{AA}=t_{BB}=t>0,\;
t_{AB}=0$. The other possible choices of the sign of $t_{AA}$ and $t_{BB}$
lead to an equivalent model \cite{note4}. At each site $i$, we introduce two
fermions $f_{i\sigma}$ and two bosons $b_{i\sigma^{\prime}}$, where $b_{i+}
\equiv e$ (empty) and $b_{i-} \equiv d$ (doublon). The fermions (bosons)
transform according to an irreducible representation of the local spin
(pseudospin) local SU(2) symmetry that $H_t$ possesses for open boundary
conditions \cite{li1}. In this representation, $H_t$ of (\ref{1}) in a ring
of $L$ sites with twists $\Phi_{\sigma}$ for particles with spin $\sigma$
reads
\begin{eqnarray}
H_t&=& \sum_{i=1}^{L-1} H_{i, i+1}\;+\;H_{L,1}\;=\; -t \sum_{i=1,\sigma
\sigma^{\prime}}^{L-1} (f^{\dagger}_{i+1 \sigma} f_{i \sigma} b^{\dagger}_{i
\sigma^{\prime}} b_{i+1 \sigma^{\prime}} + h.c.)  \nonumber \\
& & -t \sum_{\sigma} [ f^{\dagger}_{1 \sigma} f_{L \sigma} (e^{i
\Phi_{\sigma}} b^{\dagger}_{1 +} b_{L +} + e^{-i \Phi_{-\sigma}}
b^{\dagger}_{1 -} b_{L -}) + h.c. ].  \label{2}
\end{eqnarray}
The numbers $N_{\sigma}=\sum_i f^{\dagger}_{i \sigma} f_{i \sigma},\; N_e=
\sum_i e^{\dagger}_{i \sigma} e_{i \sigma}, \; N_d= \sum_i d^{\dagger}_{i
\sigma} d_{i \sigma}$ are all conserved when $t_{AB}=0$. In each subspace
with fixed $N_{\uparrow}, N_{\downarrow}, N_e, N_d$, any state has the form $%
\prod_{m=1}^{N_b} b^{\dagger}_{i(m) \sigma^{\prime}(m)} \prod_{j=1}^{N_f}
f^{\dagger}_{i(j) \sigma(j)} |0\rangle$, where $j$ labels the $%
N_f=N_{\uparrow}+N_{\downarrow}$ fermions from left to right and $i(j),
\sigma(j)$ denote the position and the spin of the $j$th fermion. Similarly $%
i(m), \sigma^{\prime}(m)$ (with $i(m+1)>i(m)$) are the position and the
pseudospin \cite{li1} of the $m$th boson. The number of bosons is $%
N_b=N_e+N_d=L-N_f$. Because of the completeness relation $\sum_{\sigma}
f^{\dagger}_{i \sigma} f_{i \sigma} + b^{\dagger}_{i \sigma} b_{i \sigma}=1$%
, $\{i(j)\}$ and $\{i(m)\}$ are complementary sets.

For the periodic case ($\Phi_{\sigma}=0$), the Hamiltonian is invariant
under cyclic permutations of the fermions and bosons and it is convenient to
work in the basis of the irreducible representation of the direct product
group ${\cal C}_{N_f} \bigotimes {\cal C}_{N_b}$ \cite{sch}. Our idea is to
use appropriate weighted representations \cite{liu} to cancel out the
difference in phases in $H_{1,L}$ in order to map the problem into one of
spinless fermions with twisted boundary conditions.  With this objective in
mind, we think the ring as a periodic system in which $f^{\dagger}_{i+L
\uparrow}=f^{\dagger}_{i \uparrow}, \; e^{\dagger}_{i+L}=  e^{\dagger}_{i}$,
but $f^{\dagger}_{i+L \downarrow}= e^{-i(\Phi_{\uparrow}-
\Phi_{\downarrow})} f^{\dagger}_{i \downarrow}, \; d^{\dagger}_{i+L}=
e^{-i(\Phi_{\uparrow}+\Phi_{\downarrow})} d^{\dagger}_i$. Using these
boundary conditions,  it can be verified that
\begin{equation}
H_{L,1}\;=\; e^{i \Phi_{\uparrow}} \sum_{\sigma \sigma^{\prime}}
f^{\dagger}_{1+L \sigma} f_{L \sigma} b^{\dagger}_{1+L \sigma^{\prime}} b_{L
\sigma^{\prime}},  \label{3}
\end{equation}
as we want.

We look for a basis of many particle states transforming as irreducible
representations of ${\cal C}_{N_f}\bigotimes {\cal C}_{N_b}$ under the above
mentioned boundary conditions. The part of these states which describes the
singly occupied sites can be constructed using the operators
\begin{eqnarray}
F^{\dagger }(\{i(j)\},\{k_l\}) &=&\prod_{\stackrel{\sim }{k}_l\uparrow
}^{N_{\uparrow }}\stackrel{\sim }{f}_{\stackrel{\sim }{k}_l\uparrow
}^{\dagger }\prod_{l=1}^{N_{\downarrow }}f_{k_l\downarrow }^{\dagger },
\nonumber \\
f_{k\downarrow }^{\dagger } &=&\frac 1{\sqrt{N_f}}%
\sum_{j=1}^{N_f}e^{-ikj}f_{i(j)\downarrow }^{\dagger },\;\;\;\;k_l=\frac{%
2\pi \nu _l-(\Phi _{\uparrow }-\Phi _{\downarrow })}{N_f}  \nonumber \\
\stackrel{\sim }{f}_{\stackrel{\sim }{k}\uparrow }^{\dagger } &=&\frac 1{%
\sqrt{N_f}}\sum_{j=1}^{N_f}e^{-i\stackrel{\sim }{k}j}f_{i(j)\uparrow
}^{\dagger }(1-f_{i(j)\downarrow }^{\dagger }f_{i(j)\downarrow }),  \label{4}
\end{eqnarray}
where in contrast to the wave numbers $k_l$, the $\stackrel{\sim }{k}_l$ are
not shifted ($\stackrel{\sim }{k}_lN_f/2\pi $ is integer), and the set of $%
k_l$ is fixed and chosen in such a way that $\sum_l\stackrel{\sim }{k}%
_l=0\;(\pi )$ for $N_f$ odd (even). The $\nu _l$ are $N_{\downarrow }$
different integers lying in the interval $[0,N_f-1]$, and each of the $%
N_f!/(N_{\uparrow }!N_{\downarrow }!)$ possible choices of the set of $\nu _l
$ define a spin configuration. It is easy to see that under cyclic
permutation ${\cal C}_{N_f}$, which carries each fermionic position to the
right ${\cal C}_{N_f}F^{\dagger }=-(-1)^{N_f}\exp (i\sum_lk_l)F^{\dagger }$.
In a similar way, using a transformation that interchanges spin and
pseudospin \cite{note1}, the pseudospin configuration can be described by an
operator $B^{\dagger }(\{i(m)\},\{{k^{\prime }}_l\})$, such that ${\cal C}%
_{N_b}B^{\dagger }=\exp (i\sum_l{k^{\prime }}_l)B^{\dagger }$, with the $N_d$
different ${k^{\prime }}_l=[2\pi \nu ^{\prime }-(\Phi _{\uparrow }+\Phi
_{\downarrow })]/N_b$. (details will be given elsewhere). The
(non-orthonormal) basis states that we use are denoted by $|\psi
\{i(j)\},\{k\},\{k^{\prime }\}\rangle =B^{\dagger }F^{\dagger }|0\rangle $.

$H_t$ permutes a fermion and a nearest-neighbor boson. The cyclic orders of
fermions and bosons are conserved. Thus, the numbers $\{k\}$ and $%
\{k^{\prime }\}$ are conserved. We drop these indices for simplicity. $%
H_{l,l+1}|\psi \{i(j)\}\rangle =0$ unless one and only one of the sites $l$
and $l+1$ is contained in $\{i(j)\}$. We restrict ourselves to these states
in the following discussion. It can be seen that for $l<L$, $H_{l,l+1}|\Psi
(\{i(j)\})\rangle =-t|\Psi (\{i^{\prime }(j)\})\rangle $, where $\{i^{\prime
}(l)\}$ differs form $\{i(j)\}$ in the position of one fermion only, which
is shifted from site $l$ to $l+1$ or conversely. If $i(N_f)=L$, then $%
H_{L,1}|\Psi (\{i(j)\})\rangle =-te^{i\Phi _{\uparrow }}{\cal C}_{N_f}{\cal C%
}_{N_b}^{-1}|\psi \{i^{\prime }(j)\}\rangle =t(-1)^{N_f}\exp
[i(\sum_{l=1}^{N_f}k_l-\sum_{l=1}^{N_b}{k^{\prime }}_l+\Phi _{\uparrow
})]|\psi \{i^{\prime }(j)\}\rangle $, where $i^{\prime }(1)=1$, and for $%
j<N_f$, $i^{\prime }(j+1)=i(j)$. When $N_{\downarrow }=N_d=0$, $H_t$ takes
the form of a problem of spinless fermions with flux $\Phi _{\uparrow }$.
The above equations show that in the general case, the problem takes the
same form, with an effective flux:
\begin{equation}
\Phi _{eff}\;=\;\Phi _{\uparrow }+\sum_{l=1}^{N_{\downarrow
}}k_l-\sum_{l=1}^{N_b}{k^{\prime }}_l=(\frac{N_{\uparrow }}{N_f}-\frac{N_d}{%
N_b})\Phi _{\uparrow }+(\frac{N_{\downarrow }}{N_f}-\frac{N_d}{N_b})\Phi
_{\downarrow }+\frac{2\pi }{N_f}\sum_{l=1}^{N_{\downarrow }}\nu _l-\frac{%
2\pi }{N_b}\sum_{l=1}^{N_d}{\nu ^{\prime }}_l.  \label{5}
\end{equation}
The energy of the system is given by
\begin{equation}
E\;=\;-2t\sum_{j=1}^{N_f}\cos (\frac{2\pi {\nu ^{\prime \prime }}_j+\Phi
_{eff}}L)+UN_d,  \label{6}
\end{equation}
where the $N_f$ integer numbers ${\nu ^{\prime \prime }}_j$ should be
different and can be chosen in the interval $[-L/2+1,L/2]$. When $N_b=\Phi
_{\uparrow }=\Phi _{\downarrow }=0$, the energy takes a similar form as that
derived by Ogata and Shiba \cite{og-shi} for the infinite $U$ Hubbard model
from the Bethe ansatz equations. Note, however, that our basis states do not
correspond to a definite total spin in general. Also, our basis is complete,
and (\ref{6}) describes the energy of any state, while the regular Bethe
ansatz states do not form a complete basis \cite{nucl}.

We are in position now to examine the ground state energy as a function of
the fluxes. For fixed $N_f=n_fL$ and total number of particles $N=nL=N_f+2N_b
$, minimization of (\ref{6}) leads to
\begin{equation}
E_g(\Phi _{\uparrow },\Phi _{\downarrow })\;=\;UN_d\;-\;2t\frac{\sin (n_f\pi
)}{\sin (\pi /L)}\cos (\varphi ),  \label{9}
\end{equation}
where $\varphi =\Phi _{eff}/L$ for $N_f$ odd and $\varphi =(\Phi _{eff}-\pi
)/L$ for $N_f$ even. The value of $U$ determines $N_d$ for the GS. For each $%
\Phi _{\uparrow },\Phi _{\downarrow }$, the numbers $N_{\uparrow
},N_{\downarrow }$, as well as $\{\nu \},\{\nu ^{\prime }\}$ should be
chosen to minimize $|\varphi |$ (module $2\pi $). It can be easily seen that
in the simplest case $\Phi _{\uparrow }=\Phi _{\downarrow }=0$ and $U=0$
that the GS is highly degenerate (many choices of quantum numbers lead to $%
\varphi /2\pi $ integer). For $U>U_c=-4t\cos (\pi n)$, it was found \cite
{li1,boe,sch} that the double occupation is forbidden in the GS ($N_d=0$),
and we recover the solution of the $U=+\infty $ Hubbard model with twisted
boundary conditions. In this case $E(\Phi _{\uparrow }+2\pi /N,\Phi
_{\downarrow }+2\pi /N)=E(\Phi _{\uparrow },\Phi _{\downarrow })$, since the
shift in $\Phi _\sigma $ can be absorbed decreasing one of the $\nu $ by $1$%
, what is always possible if $0\neq N_{\downarrow }\neq N_f$. For $\Phi
_{\uparrow }=\Phi _{\downarrow }$, this result has been obtained previously
\cite{kus}. Similarly, for the more interesting case with $N_d\neq 0$, a
change in both $\Phi _\sigma $ by $2\pi /|N_b-N_e|$ can be counterbalanced
by a change in the $\{\nu ^{\prime }\}$, leading to
\begin{equation}
E_g(\Phi _{\uparrow }+\frac{2\pi }{|L-N|},\Phi _{\downarrow }+\frac{2\pi }{%
|L-N|})\;=\;E_g(\Phi _{\uparrow },\Phi _{\downarrow }),  \label{10}
\end{equation}
for $L\neq N$, whereas for a half-filled system $E_g$ depends only on the
{\em difference} $\Phi _{\uparrow }-\Phi _{\downarrow }$, a behavior typical
of an insulator. For $U<-4\cos (\pi n),\;N_d>0$ and $\eta $-pairing states
with ODLRO are present in the GS \cite{li1,boe,sch}. However, we do not find
AFQ, but a periodicity which depends on the particle-density $n$. An example
for finite chains is shown in Figs. 1 (a-b). The number of peaks of $E_g(\Phi
_{\uparrow }=\Phi _{\downarrow }=\Phi )$ for $0\leq \Phi \leq 2\pi $ is {\em %
at least} $L|n-1|$, diverging in the thermodynamic limit, while the height
of each peak decreases as $1/L^3$, as in the $U=+\infty $ Hubbard model. The
response of the system to the flux is like that of a single particle with
charge $L-N$ or larger. One might ask whether a collection of weakly coupled
chains behaves like a superfluid of these particles. However, since the
compressibility diverges in the interesting regime \cite{li1}, charge can be
transferred between chains without cost of energy, and the response to the
flux of different chains does not add coherently. Thus, the system does not
show the Meissner effect \cite{yang,bye,kohn2}. We should also note that the
SU(2) $\eta $-symmetry which allows for the construction of eigenstates with
ODLRO is broken in the presence of a flux. It can be seen from (\ref{5})
that a twist of the form $\Phi _{\uparrow }=-\Phi _{\downarrow }$, couples
with the spin degrees of freedom in the same fashion as a twist $\Phi
_{\uparrow }=\Phi _{\downarrow }$ affects the pseudospin ones. While in the
first case, the total spin invariance is broken, remaining only $S^z$ as a
good quantum number, the total pseudospin invariance is broken in the second
case, with $\eta ^z=1/2(L-N)$ fixed by $U$ and the chemical potential. Thus,
the $\eta $-pairing states do not necessarily give rise to superconducting
currents in the presence of an external flux. The physics in the region with
$U<-4t$, where $n_f=0$ \cite{li1,boe,sch} is more obvious. In this case,
there are also $\eta $-paired states with ODLRO in the degenerate GS.
However, these states are static, and from (\ref{9}) $E(\Phi _{\uparrow
},\Phi _{\downarrow })=UN_d$ for all $\Phi _\sigma $. This result might been
anticipated from the form of $H_t$. This demonstrates that the ODLRO of the $%
\eta $-paired states {\em is not a sufficient condition for the   existence
of superconductivity}. This fact has not been noted in previous related
works.

The computations of Drude weight and the spin stiffness lead to
\begin{eqnarray}
D_c &=&\frac L2\frac{\partial ^2E(\Phi ,\Phi )}{\partial \Phi ^2}|_{\varphi
=0}\;=\;\frac t\pi (\frac{1-n}{1-n_f})^2\sin (\pi n_f)  \nonumber \\
D_s &=&\frac L2\frac{\partial ^2E_g(\Phi ,-\Phi )}{\partial \Phi ^2}%
|_{\varphi =0}\;=\;\frac t\pi (\frac{n_{\uparrow }-n_{\downarrow }}{n_f}%
)^2\sin (\pi n_f),  \label{11}
\end{eqnarray}
where $n_f$ is a function of $U/t$ and $n$ that can be obtained minimizing (%
\ref{9}) \cite{li1,sch}. For half-filling, $D_c=0,\;\forall U$. The result
is not surprising for $|U|>4t$, where the system is a Mott insulator, but
rather unexpected otherwise. The symmetry between spin and pseudospin
degrees of freedom \cite{note1}, becomes explicit by replacing $%
n_e-n_d=1-n,\;n_b=1-n_f$ in the expression of $D_c$. $D_s$ vanishes in the
sector $S^z=0$ and with it, the inverse of the magnetic susceptibility \cite
{sha}, as a consequence of the spin degeneracy. Analogously, for $|U|<4t$,
not only $D_c$ vanishes at half-filling, but also the inverse of the charge
compressibility. The system is an insulator, in spite of a vanishing charge
gap. The behavior of the Drude weight for $n\rightarrow 1$ is the same as
that of a system of $n_f$ carriers with effective mass diverging as $%
(1-n_f)^2/(1-n^2)$.

The particular behavior of the energy as a function of flux, is a
consequence of the existence of excitations associated to the fermionic
charge, spin and pseudospin degrees of freedom, which is explicit in Eqs. (%
\ref{5}) and (\ref{6}). This is similar to the case of the infinite $U$
Hubbard model where spin and charge decouple. The ground state is highly
degenerate as a consequence of the rich symmetry structure of Eq. (\ref{1})
when $|t_{AA}|-|t_{BB}|=t_{AB}=0$. In particular, for an open chain there is
a local spin and pseudospin symmetry at each site \cite{li1}. Since $t_{AB}=0
$ is an accident rather than a generic feature of any one-band model, it is
very important to discuss the effect of a finite $t_{AB}$, particularly
taking into account that this term lifts the GS degeneracy. When $t_{AB}\neq0
$, the sign of $t_{AB}$ or those of $t_{AA}$ and $t_{BB}$ simultaneously,
can be changed using symmetry properties \cite{gag,dif}, but models with
different $t_{AA} t_{BB}$ are not equivalent. In the following, we consider
the case $t_{AA}=t_{BB}=-t$, which interpolates between two exactly solvable
cases: the one considered above and the Hubbard model. This case preserves
the SU(2) pseudospin symmetry when $t_{AB}\neq0$ \cite{ali,gag}. Previous
numerical studies suggest that the 1D system for small $U$ and $n\sim 1$ is
a Tomonaga-Luttinger liquid with dominant superconducting correlations at
large distances \cite{li1,dif}. We have studied numerically $%
E(\Phi_{\uparrow}=\Phi_{\downarrow}= \Phi)$ for finite systems with fixed
densities $n\neq1,\; n_e, n_d \neq 0$ at $t_{AB}\neq 0$. An example is shown
in Fig. 1 (c). We find that a small $t_{AB}$ gives rise to AFQ in the finite
systems. The cusp near $\Phi=\pi/2$ decreases with increasing length of
chain. This behavior is typical of a system with {\em power law}
superconducting correlations at long distances rather  than a state with
true ODLRO.

What is the origin of the superconducting correlations? Is it related with
the $\eta $-pairing? To answer these questions let us begin by noting that
the (nondegenerate) GS for $t_{AB}\rightarrow 0$ is exactly known in the
case $N_eN_b=0$ (for $U>-4t\cos (\pi n)$ and $U>0$ \cite{li1,boe,sch}). In
this case, the low energy physics of the model becomes equivalent to that of
a Hubbard model with interaction $U_H=t^2U/t_{AB}^2\rightarrow \infty $ (as
can be easily seen eliminating in (\ref{2}) the term in $t_{AB}$ through the
standard canonical transformation). In this limit Ogata and Shiba \cite
{og-shi} have shown that the GS wave function can be factorized in two
parts: one describing the position and the other the spin of the $N_f$
fermions. The first factor corresponds to the GS of a Heisenberg chain with $%
N_f$ sites. This GS wave function can be mapped into the corresponding one
for $U<0$ and a magnetic field high enough to ensure $N_{\uparrow
}N_{\downarrow }=0$ using the transformation that interchanges spin and
pseudospin \cite{note1}. It is natural to expect that the effect of a small $%
t_{AB}$ is to introduce antiferromagnetic correlations between spins and
pseudospins in the general case. A straightforward generalization of the
above analyzed two cases, led us to propose an {\em ansatz} for the GS in
the limit $t_{AB}\rightarrow 0$ consisting of {\em three} factors,
describing the positions of fermions and bosons and the spin and pseudospin
variables. The first two factors are those of the GS Bethe ansatz solution
of the $U=+\infty $ Hubbard model. The last one is the GS of a Heisenberg
model for the pseudospin variables, which is also the GS of the large
negative $U$ Hubbard model in a system with $N_b$ sites and $2N_d$ particles
\cite{sof}. We have computed the overlap of our ansatz with the exact GS
obtained from exact diagonalization in different chains (up to $L=12$
sites), and we found that it is equal to $(1-\alpha ^2t_{AB}^2)^{1/2}$, with
$\alpha \sim 2-4$, for $0\neq t_{AB}<0.1$, confirming our conjecture for $%
t_{AB}\rightarrow 0$. It is easy to verify with our ansatz that in the
thermodynamic limit, for $L\rightarrow \infty $, the pair correlation
function $C(l)=\langle c_{i+l\uparrow }^{\dagger }c_{i+l\downarrow
}^{\dagger }c_{i\downarrow }c_{i\uparrow }\rangle $ can be expressed in
terms of the corresponding correlation function of the large $|U|$
attractive Hubbard model $C_H(l)$ for density $(n-n_f)/(1-n_f)$ as $%
C_H(l)=(1-n_f)[C_H(l^{\prime })]_{av}$, where the average $l^{\prime }$ is
centered around $L(1-n_f)$. Thus, the superconducting properties of the
system are essentially those of the large $U$ attractive Hubbard model with
dilute superfluid density. The superconducting properties of the model are
not related with the $\eta $-pairing. For $t_{AA}=t_{BB}$, the generators of
the total pseudospin algebra are $\eta ^{-}=\sum_i(-1)^ic_{i\uparrow
}^{\dagger }c_{i\downarrow }^{\dagger },\;\eta ^{+}=(\eta ^{-})^{\dagger }$
and $\eta ^z=(1/2)\sum_i(1-\sum_\sigma n_{i\sigma })$. The $\eta $-pairing
mechanism applies $\eta ^{-}$ to an eigenstate with $\eta \neq 0$ to obtain
eigenstates with $\eta ^z<\eta $ which possess ODLRO \cite{yan2,li1,boe,sch}%
. However, since by construction, our ansatz for the GS has $\eta ^z=\eta $,
and the true (non-degenerate) ground state has the same quantum numbers, it
cannot be the result of applying $\eta ^{-}$ to any eigenstate. Exact
diagonalization results show that this is also the case in 2D, even for $%
t_{AB}=0$ \cite{gag}.

In summary, we have shown that at least in the 1D generalized Hubbard model (%
\ref{1}) for $|t_{AA}|-|t_{BB}|=t_{AB}=0$, the $\eta $-pairing does not lead
to superconductivity. The possibility of constructing eigenstates with ODLRO
using the SU(2) symmetry does not guaranty the existence of superconducting
currents giving rise anomalous flux quantization and Meissner effect. The
ODLRO must be analyzed in the presence of a finite magnetic flux threading
the ring. This fundamental fact is in the spirit of the proposals of Refs.
\cite{yang,bye,kohn2}. We have also examined the character of the
metal-insulator transition near half filling and we have presented strong
evidence that the GS degeneracy is broken in favor of a GS with dominant
superconducting correlations in 1D when a small $t_{AB}$ is turned on.

One of us ( A. A. A.) is partially supported by CONICET, Argentina.

\newpage
{\bf Figure Captions}

{\bf 1.} Ground state energy as a function of twist angle (a) for $%
|t_{AA}|=|t_{BB}|=1, \; t_{AB}=0$, density $n=2/3$ and $U=0$; (b)
same as (a) with $U >4$;
(c) for  $t_{AA}=t_{BB}=-1, \;t_{AB}=-0.2, \;U=0, \;n=2/3$


\begin{references}
\bibitem{note}  Systems with pairing but not superconductivity also display
AFQ.
\bibitem{note0}  The Hamiltonian (\ref{1}) has been obtained as an effective
model to describe the low energy physics of intermediate valence systems,
high-$T_c$ and other superconductors \cite{one} and according to several
studies, small differences in the hopping parameters can lead to
superconductivity for $U>0$ \cite{dif}.
\bibitem{note1}  For $|t_{AA}|=|t_{BB}|$ and any $t_{AB}$, the kinetic part
of the Hamiltonian $H_t$ is invariant under the change of representation $%
f^{\dagger}_{i \sigma}|0 \rangle \leftrightarrow [-{\rm sgn}%
(t_{AA}t_{BB})]^i b^{\dagger}_{i \sigma}|0\rangle$, while $H_U$ changes sign.
\end{references}
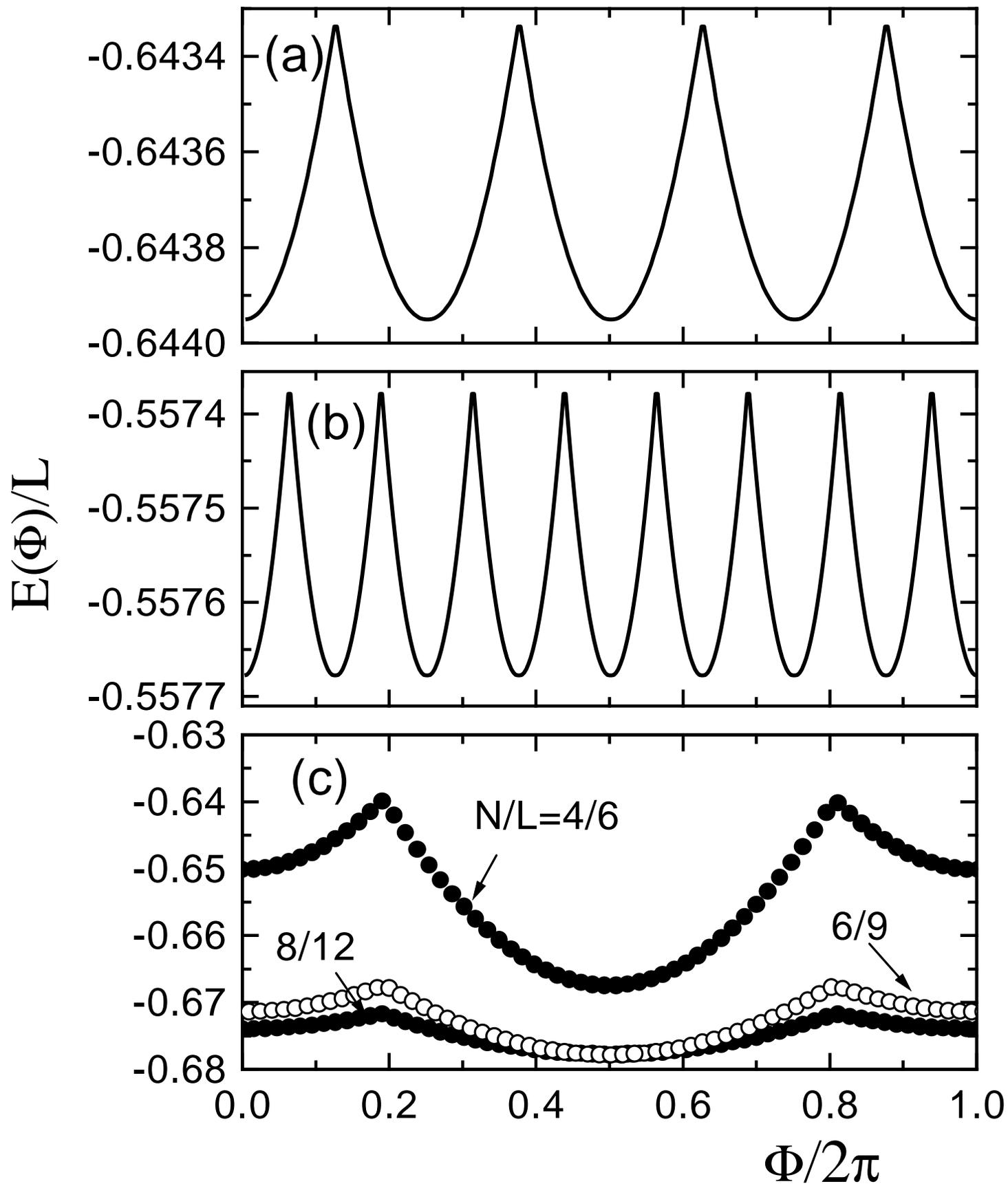
\end{document}